\documentclass{aastex}
\usepackage{amsmath,amssymb,graphicx}
\begin{document}
\sloppy
\title{On The Sources of Fast Radio Bursts}
\author{J. I. Katz}
\affil{Department of Physics and McDonnell Center for the Space Sciences\\
Washington University, St. Louis, Mo. 63130}
\email{katz@wuphys.wustl.edu}
\begin{abstract}
	{This paper argues that} repeating and apparently non-repeating
	Fast Radio Bursts are distinct classes of events produced by
	distinct classes of sources.  I review evidence for that division,
	and discuss the statistics {of these classes.  They differ in
	temporal/spectral space, spectral/duration space and rotation
	measure; the first two differences indicate different emission
	processes, and the last indicates different environments.  I discuss
	two of the many} models of each class of source: black hole
	accretion discs for repeating FRB and hypermagnetized neutron stars
	({some observed to produce} SGR) for apparently non-repeating
	FRB.  Appendices suggest {low} cutoff frequencies of coherent
	emission that are consistent with these models and with known 
	{pulsars}, and discuss the necessary conditions for acceleration of
	energetic particles.
\end{abstract}
\section{Introduction}
When a category of phenomena consists of two or more distinct
sub-categories, understanding requires separating those sub-categories.  A
model that explains one sub-category may be {apparently} contradicted by
data from another sub-category.  {Recognizing the existence of two or
more distinct sub-categories means that a model that explains one
sub-category will not be rejected because it fails elsewhere.}  Historic
examples \citep{M21} include the distinction between nov\ae\ and
supernov\ae, between Gamma Ray Bursts (GRB) and Soft Gamma Repeaters (SGR)
and between Populations I and II Cepheids.  In each case, very distinct
events were conflated until further data (SN 1885 in the Andromeda galaxy
and its recognition as a distant external galaxy, repetitions and spectral
data of SGR 0525$-$66, the differing compositions of Populations I and II
stars) established the distinction.

Fast Radio Burst (FRB) sources are {phenomenologically} divided into
repeating and apparently non-repeating classes.  However, this distinction
{might not be} meaningful:  Every repeater has a first detected burst,
after which it is an apparent non-repeater, until a second burst is
observed.  {In \S2 I argue that the distinction is meaningful:
the two phenomenologically defined classes differ in several ways other
than repetition rate, and observational constraints on the repetition rates
of the few extensively-monitored apparent non-repeaters are several orders
of magnitude less than the observed repetition rates of well-studied known
repeaters.  This difference will either disappear or rapidly (quadratically
in time) grow as the CHIME/FRB database accumulates.  This is consistent
with the suggestion \citep{J23,Y24} that the majority of apparently
non-repeating FRB may be repeaters because the reported low duty cycles
only apply to the very few apparent non-repeaters that have been the
subjects of extended observation.}

If the distinction is only quantitative, qualitatively similar sources
differ in repetition rate, and apparent non-repeaters eventually repeat.
If it is qualitative, the two categories are fundamentally different and
require different explanations.  {\it N.~B.:\/} ``Apparently
non-repeating'' is defined phenomenologically, as the absence of repetitions
in extant data; it need not imply never repeating, only that bursts are
infrequent.  {For example, some repeaters have had burst rates of one
per minute, while some apparent non-repeaters have been inactive over
hundreds of hours, implying duty cycles differing by several orders of
magnitude.}

\citet{R19a,H20} {argued} that the rate of catastrophic
events that destroy their source objects (supernov\ae, merging neutron
stars, {\it etc.\/}) is insufficient to account for the observed rate of
apparently non-repeating FRB, implying that they must repeat at some low
rate {or that only a few percent of apparent non-repeaters are true
one-offs from cataclysmic events.  \citet{J22,ZZ22} have questioned the
assumption that the volumetric occurrence rate of apparently non-repeating
FRB was almost constant during the past}.

\S2 summarizes the arguments that repeating and apparently non-repeating FRB
are qualitatively different, and that different sources must be sought for
them.  \S3 proposes that repeating FRB are produced by accretion discs
around black holes, likely of intermediate (between stellar and 
{supermassive (Active Galactic Nucleus))} mass.
\S4 argues for the widely held hypothesis that apparently non-repeating FRB
are produced by ``magnetars'', hypermagnetized neutron stars whose emission
is powered by their magnetostatic energy.  The arguments presented here are
phenomenological, rather than based on theoretical models of FRB emission, a
much more difficult problem.  Appendix \ref{cutoff} discusses a cutoff
frequency on propagation of radiation in a magnetosphere while Appendix
\ref{conditions} discusses a necessary condition for the acceleration of
energetic particles.
\section{Two Distinct Classes}
There are several ways in which repeaters differ from apparent
non-repeaters.
\subsection{Duty Cycle}
A duty cycle $D$ may be defined, as a generalization of its definition in
engineering in which it is the fraction of the time a system operates (for
example, a radar transmitting a $1\,\mu$s pulse 1000 times a second has a
duty cycle of 0.001):
\begin{equation}
	D \equiv {\langle F \rangle^2 \over \langle F^2 \rangle},
\end{equation}
where $F$ is the flux.  For a source that emits at a constant flux a
fraction $D$ of the time this is equivalent to the engineering definition
of $D$, but generalizes it to the astronomical case of a source whose
intensity varies.

The {first \citep{S16} and likely} best-studied repeating FRB 20121102A
{(the literature is extensive, but \citet{Cr21} is particularly
thorough)} has $D \sim 10^{-5}$, as may be estimated from the observation
{by the very sensitive FAST} of $\sim 1\,$ms pulses with typical
intervals $\sim 100\,$s \citep{L21}.  In contrast, the best-observed
apparently non-repeating FRB have $D \lesssim 10^{-8}\text{--}10^{-10}$
\citep{K19}.  {More recent studies \citep{CHIME/FRB23} are consistent
with this.  A telescope less sensitive than FAST would have observed a lower
repetition rate than did \citet{L21}, but still implying $D$ orders of
magnitude greater than the upper limits on apparent non-repeaters.}

The large number of sources (about 500 per year) observed by CHIME/FRB and
its observing duty cycle $d \approx 0.02$ (30 minutes/day) imply either
that in the fairly near future (1--10 years) the upper bound {on the
repetition rate of apparent non-repeaters} will be reduced by orders of
magnitude to
\begin{equation}
	R \lesssim {2 \over NTd} \approx 0.2
	\left({T \over \text{y}}\right)^{-2} \text{y}^{-1},
\end{equation}
where $T$ is the duration of observation and $N \approx 500$(T/$y$) is the
number of {apparently non-repeating FRB} observed, or that repetitions
of previously apparently non-repeating FRB will be observed.

For the recent CHIME/FRB catalogue \citep{CHIME/FRB}, $T \approx 1\,$y and
$R \lesssim 0.2/\text{y}$.  With a nominal FRB width $\Delta t = 3\,$ms, the
implied bound on the duty cycles of apparently non-repeating FRB
\begin{equation}
	D \sim R \Delta t \lesssim 2 \times 10^{-11},
\end{equation}
several orders of magnitude less than the duty cycles of known repeating
FRB.  Even if repetitions are observed, their implied duty cycles {are
often} orders of magnitude less than those of known repeaters, {although
there is some overlap}.  This distinguishes {the majority of} repeating
FRB from {the majority of} apparently non-repeating FRB and implies
different sources and mechanisms.
\subsection{Spectro-Temporal Differences}
\begin{figure}
                \centering
                \includegraphics[width=6.5in]{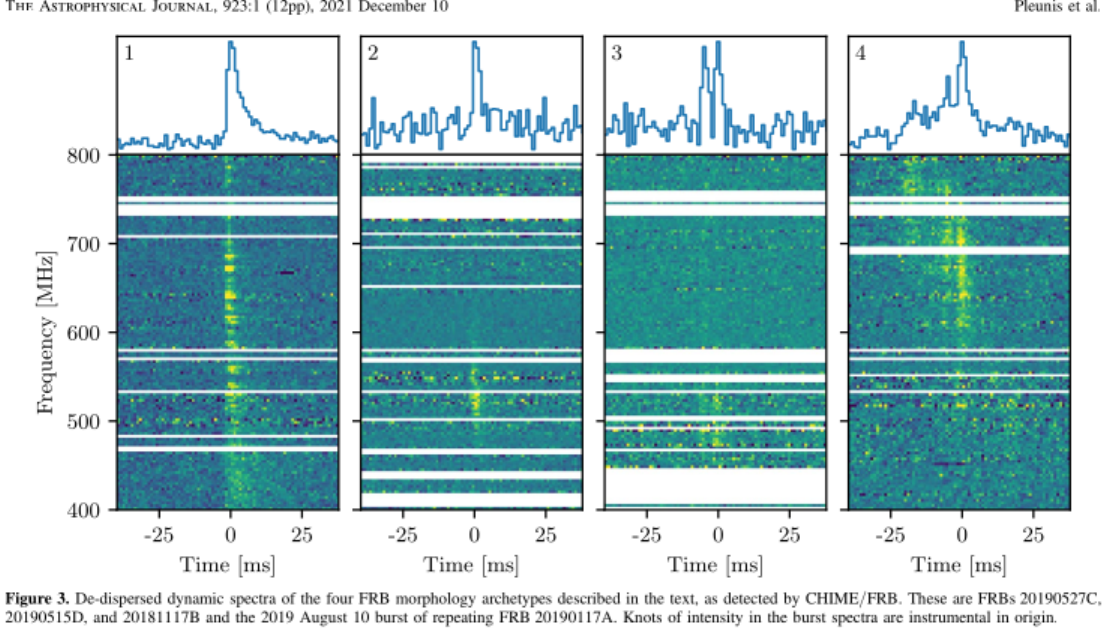}
		\caption{\label{P3} De-dispersed dynamic spectra of
		bursts of four types, observed and discussed by \citet{P21b}.
		The first three sub-figures are apparent non-repeaters.
		The fourth is a repeater, and shows a drift to lower
		frequency, distinct from dispersion.  This ``sad trombone''
		effect is found for most repeating bursts, {but is
		unusual for apparent non-repeaters \citep{B23}}.} 
        \end{figure}

\begin{figure}
                \centering
                \includegraphics[width=6.5in]{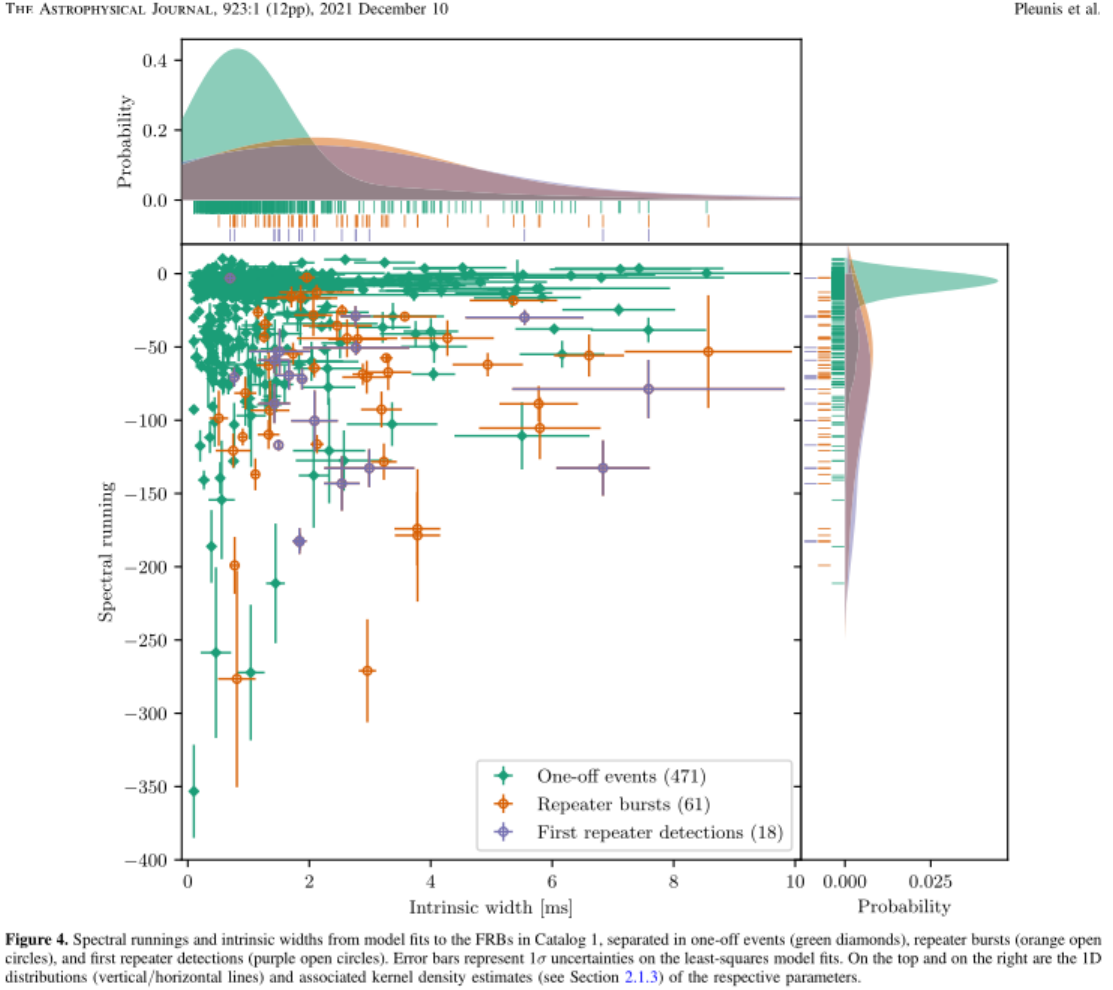}
		\caption{\label{P4} Fitted spectral parameter ``running''
		{\it vs.\/} burst length for several hundred bursts (numbers
		 in figure) \citep{P21b}, with apparent non-repeaters
		(``one-offs'') green, repeaters purple and orange.  The
		 distributions of each variable, separately, overlap, but
		 in the two-dimensional space there is a clear separation.}
        \end{figure}

\begin{figure}
                \centering
                \includegraphics[width=6.5in]{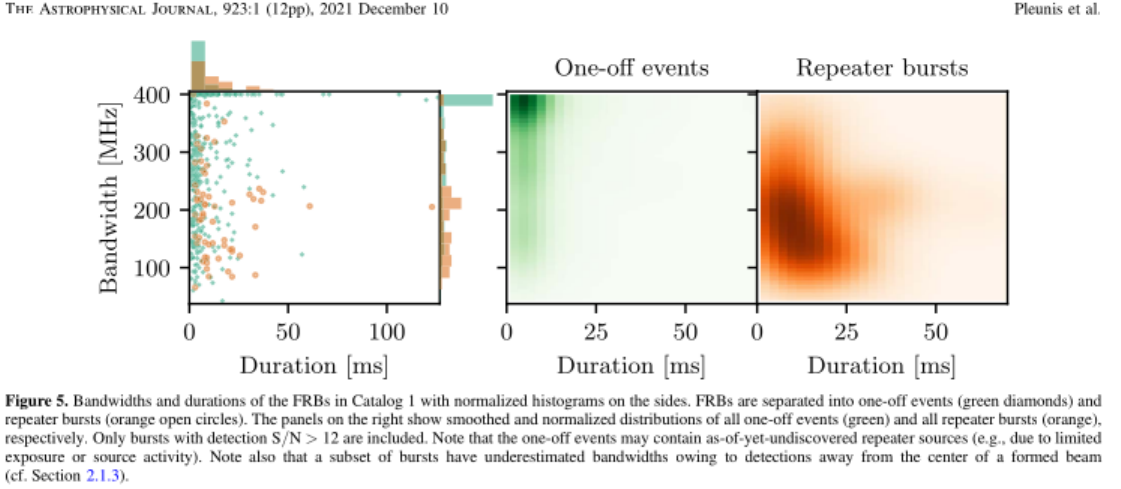}
		\caption{\label{P5} Burst bandwidth {\it vs.\/} durations
		\citep{P21b}, with apparent non-repeaters (``one-offs'')
		green, repeaters purple and orange.  The distributions in
		the two-dimensional space are nearly disjoint.}
        \end{figure}

The spectral and temporal behavior of repeating and apparently non-repeating
FRB differ {qualitatively \citep{H19,P21b}.  For example, Fig.~\ref{P3}
\citep{P21b} shows the ``sad trombone'' observed for repeaters but not for
apparent non-repeaters.  Fig.~\ref{P4} \citep{P21b} shows the distributions
of spectral running {\it vs.\/} burst width for repeating and apparently
non-repeating FRB.  Although as a function of a single variable these
overlap, in two-dimensional space they are separated.  Fig.~\ref{P5}
\citep{P21b} shows the distributions of burst bandwidth {\it vs.\/}
duration.  Again, the distributions are well separated in this
two-dimensional space.  The fact that in the two dimensional spaces the
distributions of apparent non-repeaters and known repeaters are separated
with only limited overlap indicates they are different classes of objects;
only a small fraction of the apparent non-repeaters in these data have the
properties of known repeaters.}
\subsection{Polarization}
Repeating FRB have polarization that increases with increasing frequency.
This is in contrast to pulsars, whose polarization typically is a decreasing
function of frequency \citep{F22}.  {Comparison to pulsars} argues
against models for repeaters in which their bursts are produced {in the
same region as pulsar radiation,} within neutron star magnetospheres, 
{although the interpretation of these observations might be confused by
propagation effects \citep{B22,P22}.  No such systematic trend in
polarization has been reported for apparent non-repeating FRB, allowing for
possible magnetospheric origin.}

It does not argue against FRB models in which the energy source is
a neutron star but emission occurs far away by mechanisms different from
those that radiate pulsar pulses \citep{S21}.  In such models the energy
source {might} be either rotational (a pulsar) or magnetostatic (a
magnetar).

\begin{figure}
	\centering
	\includegraphics[width=5in]{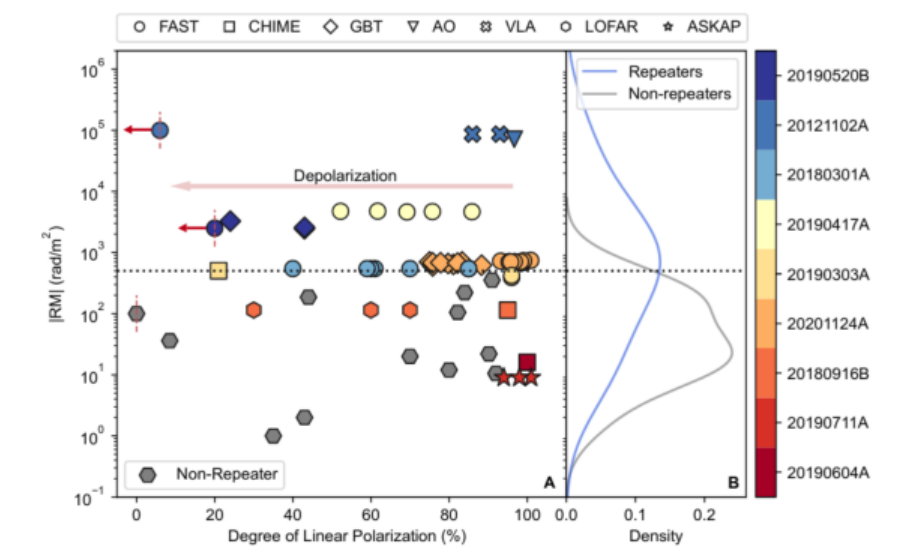}
	\caption{\label{F3} Repeaters have, on average, larger rotation
	measure (RM) than apparent non-repeaters \citep{F22}.}
\end{figure}

Some of the differences between repeaters and apparent non-repeaters are
shown in Fig.~\ref{F3}.
\section{Sources of Repeaters and The Magnetar Hypothesis}
The most popular models of FRB {(see \cite{Z23} for an extensive
review)} involve ``magnetars'', neutron stars with
extraordinary ($10^{14}$--$10^{15}$ Gauss) magnetic fields whose
magnetostatic energy was proposed \citep{K82,TD92} to power Soft Gamma
Repeaters (SGR).  Their existence was demonstrated by the discovery of
their quiescent counterparts, Anomalous X-Ray Pulsars (AXP), whose rapid
spindown demonstrated the existence of the large magnetic fields required to
power SGR outbursts. 
\subsection{Giant outburst of SGR 1806-20 did not make a FRB}
A fortuitous out-of-beam radio observation \citep{TKP16} of the giant
outburst of SGR 1806$-$20 did not observe a FRB.  At its Galactic distance
(about 300,000 times closer than a ``cosmological'' FRB at $z \sim 1$) and
with 60 dB sidelobe suppression, a ``cosmological'' FRB would have been
about 50 dB brighter than at $z \sim 1$.  Yet no FRB was observed, 
{implying either that FRB in general are beamed and observed if we are
fortuitously in the beam or that SGR 1806$-$20 did not emit a FRB.
\citet{TKP16} set an upper bound on the fluence ratio ${\cal F}_\text{FRB}/
{\cal F}_\gamma \le 10^7\,$Jy-ms/(erg/cm$^2$)}.

{In contrast, the atypical Galactic FRB 20200428 was produced by the
magnetar SGR 1935$+$2154.  Its fluence ratio was ${\cal F}_\text{FRB}/
{\cal F}_\gamma \approx 2 \times 10^{12}\,$Jy-ms/(erg/cm$^2$)
\citep{B20a,HXMT20}, more than five orders of magnitude greater than the
upper bound on this ratio for SGR 1806$-$20.  This established that these
events are of different classes; either SGR 1935$+$2154 is not just a
lower energy version of SGR 1806$-$20 or FRB 20200428 was not just a scaled
down (but local) version of a ``classical'' FRB at cosmological distance.} 
\subsection{Repeaters in compact dense plasma environments}
{The environment of} FRB 20121102A has a multi-milligauss magnetic field
\citep{M18,K21a} and the dispersion measure (DM) of FRB 20190520B varies
rapidly \citep{AT23} over a range of 40 pc-cm$^{-3}$, implying an electron
density $n_e \sim 10^9\,$cm$^{-3}$ \citep{K22b}.  However, known magnetars
are in supernova remnants, swept clear of dense plasma.  {SGR
1935$+$2154 is an example, and its associated FRB 20200428 shows no evidence
of a dense magnetoionic plasma environment.}  Observations of pulsars in
supernova remnants do not show the rapid variations of DM and RM observed in
some repeating FRB.
\subsection{Absence of rotational periodicity}
Periodicity is expected if the energy source is a rotating magnetic neutron
star.  This applies whether the radiation is emitted in the magnetosphere or
in a surrounding supernova remnant excited by a narrowly collimated beam.
In order to produce a pulse of width $\Delta t$ at a distance $R$ from the
source of the beam requires collimation within an angle $\Delta\theta
\lesssim \sqrt{\Delta t/Rc}$ if the beam is directed to the observer, and
$\Delta\theta \lesssim \Delta t/Rc$ if it is oblique (less likely, because
then the radiation would not be collimated in the observer's direction).

Rotating Radio Transients (RRAT) are pulsars most of whose pulses are
nulled; RRAT are natural models for the temporal behavior of FRB, if FRB are
produced by rotating neutron stars.  The periodicity of RRAT is demonstrated
by showing that their pulses are separated by integer multiples of an
underlying period.  No such period has ever been found for repeating FRB.
The best data are those of \citet{G18}, who observed five bursts from FRB
20121102A within 93 s, giving four independent constraints on any such period.
Confinement to such a short interval essentially eliminates the effects of
any possible period derivative and of uncertainty in the times of the
bursts.  Attempts to find such a period have failed, in this object and in
other repeating FRB.

\begin{figure}
        \centering
        \includegraphics[width=6in]{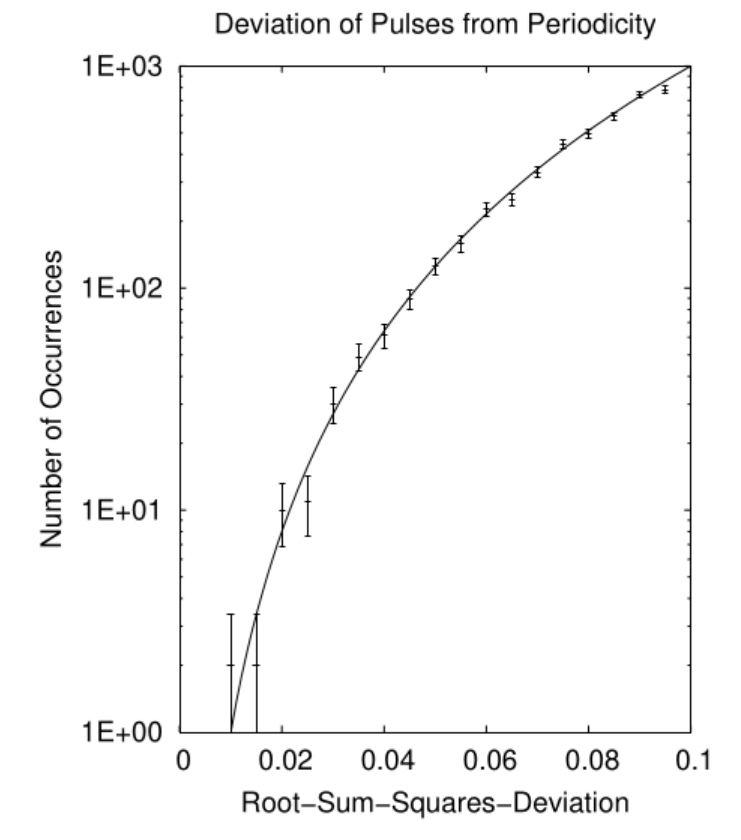}
        \caption{\label{FOMfig}Distribution of r.m.s.~deviation from exact
	periodicity for hypothesized periods for five FRB 20121102A bursts in
	93 s \citep{G18}; Error bars are $\pm 1\sigma$.  The smooth curve is
	expected for uncorrelated aperiodic bursts (shot noise).  If the
	bursts were periodic that period would have had zero r.m.s.
	deviation; no such period is found.}
\end{figure}

A more systematic study of these data \citep{K22a} calculated a figure of
merit
\begin{equation}
        \text{FOM}(P) = \sum_{i=\text{C,D,E,F}}
        \left({(t_i - t_\text{B})-\text{NINT}[(t_i - t_\text{B})/P]P
        \over P}\right)^2,
        \end{equation}
where the bursts are denoted B, C, D, E and F, for the fit of the burst
times to periods $P$, for $10^7$ periods from 0.432 ms to 1.2 h, evenly
spaced in frequency from $231\,\mu$Hz to $2.31\,$kHz.  The results are
shown in Fig.~\ref{FOMfig}.

The data of \citet{L21}, 1652 bursts of FRB 20121102A,
may be searched for periodicity using periodograms.  These bursts were
observed over several months, over which time even a very small frequency
drift (neutron star spindown) would dephase a periodogram by
\begin{equation}
                \Delta \phi = {1 \over 8} {\dot \omega} T^2,
        \end{equation}
where $T$ is the length of the data run.  Hence it is necessary to analyze
each separate data run, generally about one hour long, separately, reducing
$T$ from months to about a hour, and $\Delta \phi$ by nearly seven orders of
magnitude.  An underlying periodicity would appear in each run, though
possibly at slightly differing periods, even though it would be washed out
in a periodogram of the entire dataset.  These periods could be separated by
less than their individual uncertainties.

\begin{figure}
        \centering
        \includegraphics[width=6.5in]{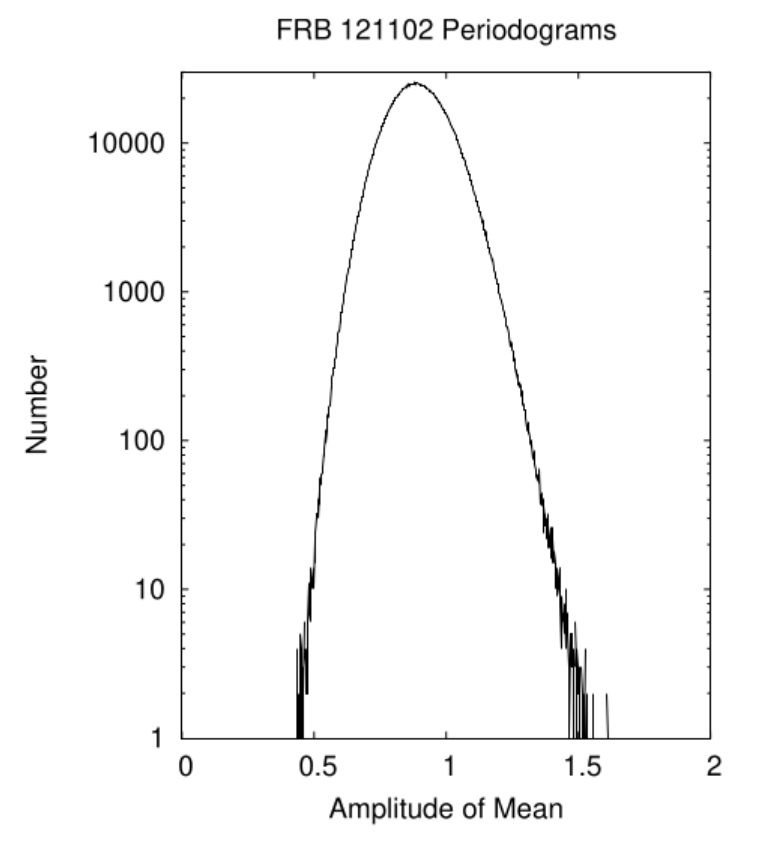}
        \caption{\label{121102mean} Distribution of the the $3.6 \times
        10^6$ averages of the normalized amplitudes of 17 single-session
	($\ge 50$ bursts) periodograms of FRB 20121102A for periods from 1 ms
	to 1 hour, evenly spaced in frequency; data from \cite{L21}.}
\end{figure}

The dataset of \citet{L21} includes 17 separate runs with at least 50 bursts
each.  These are analyzed and the distribution of periodogram amplitudes
(summed over the 17 periodograms) shown in Fig.~\ref{121102mean}.
	The distribution is close to the Gaussian expected for ``shot
	noise'' burst times, somewhat broadened and skewed by the slow
	variations in activity of FRB 20121102A.  A period in even one
	periodogram of the 17 would give an outlying amplitude and a
	persistent period would give a cluster of outliers, even though the
	individual periodicities would be dephased.  The largest average
	normalized amplitude in the data is 1.706, consistent with shot
	noise.

	Periodograms of 1--2 hour periods of intense activity, each
	comparable to a single observing session of FRB 20121102A
	\citep{L21}, of FRB 20180916B \citep{Mck23a} {and a thorough
	study of more than 800 bursts from FRB 20201124A that considered
	possible source acceleration (as would be produced by a binary
	orbit) as well as fine substructure \citep{Niu22} also showed no
	evidence of periodicity}.
\subsection{Dense chaotic environments}
	The dispersion measures of FRB include contributions from our
	Galactic disc, the Galactic halo, the intergalactic medium, {any
	galactic haloes along the line of sight \citep{S23},} the
	host galaxy and the vicinity of the FRB source.  The Galactic
	contributions can be estimated, but with substantial uncertainty.
	When the FRB source is identified with a galaxy of measured redshift
	the intergalactic contribution can be estimated with confidence.
	The host galaxy's contribution is necessarily uncertain, but is
	plausibly of the same order as the Galactic contribution.  {At
	least one repeater (FRB 20190520B \citep{Niu21}) and at least one
	apparent non-repeater (FRB 20220610A \citep{R23}) have large excess
	DM, but it is hard to distinguish between near-source and
	intergalactic origin.}

	The environments of {some repeaters are extraordinary in other
	ways.  The rest (source)-frame rotation measure (RM) of FRB 20190520B
	changed from about $-11000$ rad/m$^2$ to $+19000$ rad/m$^2$ over
	about 270 days, and then decreased to about $-37000$ rad/m$^2$ in
	the subsequent 150 days \citep{AT23},} indicating a reordering of a
	compact magnetized environment.  The repeater FRB 20121102A had even
	larger RM $\sim 100000\,$rad/m$^2$, decreasing by about 30000
	rad/m$^2$ over three years \citep{M18,W20,H21}, and indicating (when
	combined with changes in DM) magnetic fields in the range 3--17
	milligauss \citep{K21a}.  A study \citep{Mck23b} of 12 repeating FRB
	found variations in their (smaller) RM on time scales of months,
	again indicating chaotic magnetoionic environments unlike those
	observed around pulsars.

\begin{figure}
		\centering
		\includegraphics[width=6.5in]{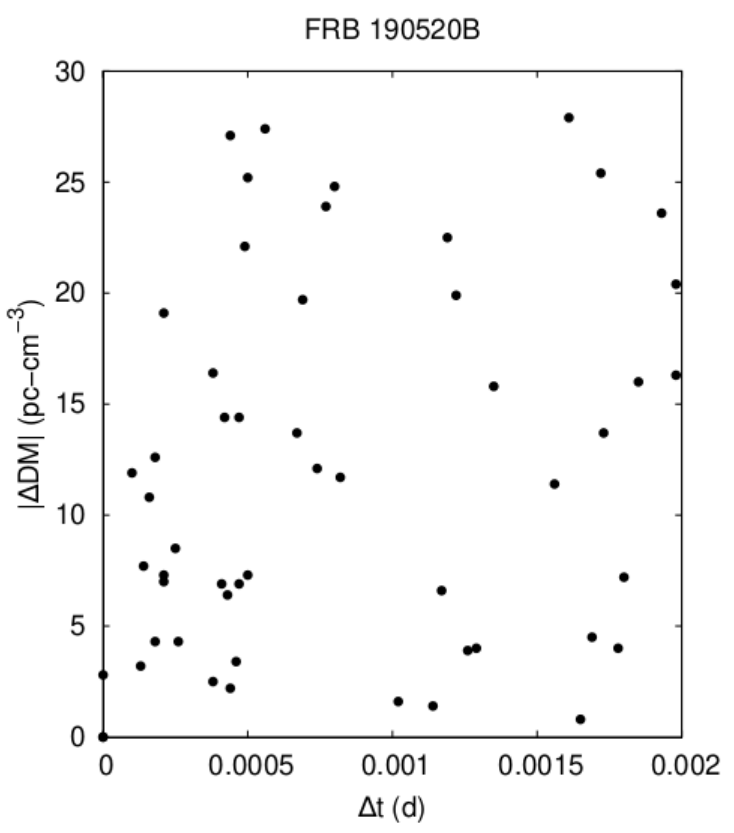}
		\caption{\label{190520B} $|\Delta\text{DM}|$ {\it vs.\/}
		$\Delta t$ for intervals between successive bursts of FRB
		20190520B on MJD 59373; {data from Table S3 of}
		\citet{AT23}.}
	\end{figure}

	Even more remarkable are the observations of the DM of the repeating
	FRB 20190520B, that varied over a range of about 30 pc-cm$^{-3}$ on
	time scales of tens of s, as shown in Fig.~\ref{190520B} {(data
	from \citet{AT23})}.  A straightforward interpretation \citep{K22b}
	in terms of clouds of plasma moving in and out of the line of sight
	suggests electron densities as high as $n_e \sim 10^9$/cm$^3$.  
	{The requirement that a cloud with DM of 30 pc-cm$^{-3}$ and this 
	$n_e$ be transparent at an observation frequency of 1.4 GHz implies
	that its temperature $T \gtrsim 3 \times 10^6\,$K.  This is
	plausible for the near ($\text{DM}/n_e \sim (10^{11} \text{cm}$)
	environment of an accreting, perhaps intermediate mass, black hole.}

	Variations in DM and RM for repeaters obviously cannot be directly
	compared to the DM and RM of apparent non-repeaters, which are
	one-off observations.  But the varying DM of FRB 20190520B and
	RM of {the repeating FRB 20121102A are inconsistent with the
	evacuated interiors of the supernova remnants that surround soft
	gamma repeaters.  The extraordinarily large RM of FRB 20121102A,
	even aside from its variation, is inconsistent with the observed RM
	of the apparent non-repeaters \citep{P24}.}
\subsection{If Magnetars Don't Make Repeating FRB, Then What Might?}
\begin{enumerate}
	\item Must not be periodic!
	\item Must not be catastrophic!
	\item Must be rare
	\item Must {be consistent with a} dense chaotic
		{magnetoionic} environment
\end{enumerate}

What satisfies these criteria?

Some black hole accretion discs ({\it e.g.\/} AGN, SS433 {\it etc.\/})
accelerate jets of energetic particles.  Their lifetime is as long as there
is a supply of matter to accrete, perhaps from a companion star, which may
be the lifetime of that star.  Accretion funnels are a natural collimating
mechanism.  However, they are not rare; there are many accreting black holes
in our galaxy.  Worse, Galactic binary black hole accretion discs are not
observed to make coherent radiation, much less FRB.

If the hypothesis \citep{K17} that accretion discs around black holes make
FRB is correct, the accretion discs that make FRB must differ in some manner
from those of the much more numerous Galactic black hole X-ray binaries.
One possible difference is as simple as orientation: their coherent
radiation (including hypothetical FRB) may be very narrowly collimated along
the disc axis.  {Appendix \ref{cutoff} discusses magnetospheric plasma
cutoffs.  If collimation is} narrow enough, this would have attractive
consequences:
\begin{itemize}
	\item It might explain their rarity
	\item FRB luminosities would be several orders of magnitude less
		than their (forbidding) isotropic-equivalent luminosities
	\item Lifetimes could be as long as binary mass-transfer lifetimes
		(the decade of activity of FRB 20121102A without evident
		systematic change would not be a surprise)
	\item The environment of a mass transfer binary might explain the
		large and variable DM and RM observed in some repeating FRB
	\item FRB would then be identified as suitably oriented
		microquasars, whose orientation would be consistent with the
		absence of the double radio lobes of most microquasars.
\end{itemize}
\cite{S21} suggested that FRB are distinguished from other black hole
accretion discs by extraordinarily high mass transfer rates, resembling
those of Ultra-Luminous X-ray sources (ULX), {although at least some ULX
are produced by neutron stars rather than black holes}.  This would
contribute to the rarity {of FRB without implying or requiring a
similarity between the thermal emission of ULX and the coherent non-thermal
emission of FRB}.
\subsection{Precession}
\begin{figure}
        \centering
        \includegraphics[width=6.5in]{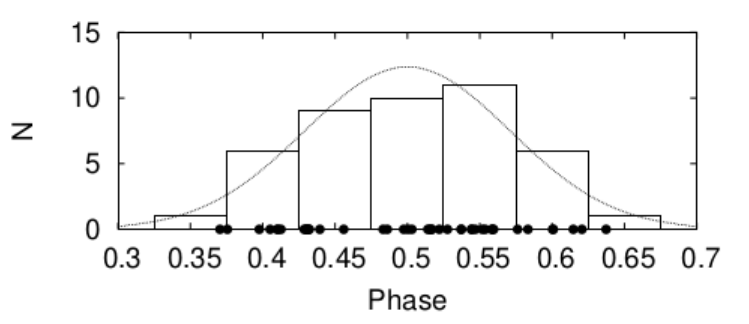}
        \caption{\label{jitter} Phases of the 44 individual bursts of FRB
	20180916B observed by \citet{Mck23a} for the fitted period of 16.315
	d (points), a bar graph of their distribution (solid) and the fitted
	Gaussian (dotted).  If the phase jitter results from a random walk
	about exact periodicity, the observed Gaussian distribution is
	expected.  The mean phase is defined as 0.5 and its standard
	deviation is 0.071.}
\end{figure}

Although individual outbursts of repeating FRB do not show the rapid
periodicities characteristic of rotating neutron stars, the activity of two
repeating FRB (FRB 20180916B \citep{CHIME20,P21a} and FRB 20121102A
\citep{R20} is modulated with long periods (16.35 d and 160 d,
respectively).  The times of the individual bursts are not periodic, but
the bursts occur within periodic windows.  There is significant phase
scatter about exact periodicity, shown in Fig.~\ref{jitter}, although the
underlying period is apparently stable.  {The width of the active
window depends somewhat on the frequency of observation \citep{P21a}.}

A number of explanations have been suggested for this periodicity.  The
phase jitter resembles that of the beams of SS 433 that jitter in angle.
SS 433 is a mass transfer binary containing a $15 M_\odot$ black hole that
accelerates beams along the angular momentum axis of its precessing
accretion disc.  The close analogy between the temporal behavior of FRB
180916B and that of SS 433 supports the black hole accretion disc model of
repeating FRB \citep{K22c}.
\section{Apparently Non-repeating FRB}
\subsection{Constraints from repetition rate}
The rate of apparently non-repeating FRB exceeds that of all known
catastrophic events \citep{R19a,H20}.  {Most or all} apparently
non-repeating FRB must repeat, but so infrequently that no such repetitions
have been observed.

There are constraints on the repetition rates of a few specific apparent
non-repeaters:
\begin{itemize}
	\item FRB 20190523 did not repeat in 78 hours \citep{R19b}, setting
		an upper limit on its repetition rate $\lesssim 0.3/d$.
        \item FRB 20180924 did not repeat in 720 hours of less
                        sensitive observations nor in 11 hours of more
			sensitive observations \citep{B19}; adjusting for
			sensitivity suggests repetition times $\gtrsim
                        10^3\text{--}10^4\,$h.
		\item {FRB 20210117A did not repeat in about 140 hours
			of subsequent observations, mostly by ASKAP with a
			detection fluence threshold about eight times lower
			than its originally detected fluence \citep{B23}.}
		\item {A statistical study of 27 apparent non-repeaters
			over 383.2 hours found only one to repeat, and set
			statistical limits on any repetition behavior
			\citep{J20}.}
\end{itemize}

These constraints {have been} much tightened by CHIME/FRB
\citep{CHIME/FRB23} and {will continue to be tightened}, or repetitions
of apparent non-repeaters will be observed.  CHIME/FRB stares at everything
in the northern sky about 30 minutes per day (an observational duty cycle
$D \approx 0.02$), every day, and observes $N \approx 500$ sources per year.
After a time $T$, if no repetitions are observed the repetition rate would
be constrained:
\begin{equation}
R \lesssim {1 \over NDT} \sim {0.1 \over \text{y}}
\left({T \over \text{y}}\right)^{-2}.
\end{equation}
Either a repetition will be observed or this bound will tighten rapidly.
\subsection{Candidate Sources: Soft Gamma Repeaters (``Magnetars'')}
These have been popular candidates as FRB sources since their discovery for
several reasons:
\begin{enumerate}
	\item SGR are transients.
	\item SGR giant outburst rise times are observed to be sub-ms.
	\item SGR have plenty of energy (observed up to $10^{47}$ ergs).
	\item Rare repetitions of giant SGR outbursts might explain the FRB
		event rate (giant outbursts guesstimated  to occur $\sim
		100$ times in a SGR lifetime).
	\item Observed intervals $> 30\,$y between giant SGR repetitions
		are consistent with apparent non-repetition of FRB.
\end{enumerate}
\subsection{FRB 20200428/SGR 1935$+$2154}
This is the only confirmed \citep{B20a,CHIME20} FRB/SGR association,
although the ``giant'' (1.5 MJy-ms) FRB 20200428 at 6 kpc was too weak to
have been detected at the distances $\gtrsim 10\,$Mpc of all but one other
FRB.  Three much weaker repetitions of FRB 20200428 were detected
\citep{Ki21}, but at distances $\gtrsim 100\,$kpc these would have been
undetectable.  At distances 100 kpc--10 Mpc FRB 20200428 would have been an
apparent non-repeater.  {Subsequent spin glitches and an additional
burst have been reported \citep{H24}.}
\subsection{Problems with the Magnetar Model}
\begin{enumerate}
	\item No FRB was associated with the giant outburst of SGR
		1806$-$20.  A fortuitous simultaneous radio observation
		\citep{TKP16} set an upper bound about 50 dB lower than a
		``cosmological'' FRB at the distance of SGR 1806$-$20
		(110 dB from its closer distance minus 60 dB suppression of
		sensitivity at $35^\circ$ from the main beam).
	\item FRB are coherent emission by bunched energetic charges, while
		SGR are thermal phenomena with the spectrum of an
		equilibrium pair gas.
	\item The dense plasma inferred from the spectrum of SGR is opaque
		to all electromagnetic radiation.
	\item SGR outburst durations are $\sim 100\,$ms, much longer than
		the 0.1--10 ms durations of most FRB, {(although
		\citet{M20} reported narrower substructure in a burst from
		SGR 1935+2154)}.  However, this might be explained by the
		sub-ms rise times of giant SGR \citep{K16}.
\end{enumerate}
\subsection{Statistics of Apparently Non-Repeating FRB and SGR}
The giant outbursts of SGR and apparent non-repeating FRB are both outliers
from their distributions of lesser events \citep{K21b}.  This is unusual in
astronomy: Most distributions of astronomical parameters, including flux and
fluence but also others, are well fit by (sometimes broken) power laws 
{({\it e.g.\/} FRB 20201124A \citep{Ki24})}.  Apparently non-repeating FRB
and SGR are exceptions.

This can be parametrized by the ratios of the most extreme value of a
parameter $x_1$ to the next-most extreme ($x_1/x_2)$.  Table \ref{outliers1}
shows some examples; Confidence is the confidence with which the hypothesis
that $x_1$ is consistent with the power law fitted to lesser events or
objects can be rejected.
\begin{table}
        \begin{center}
        \begin{tabular}{|crccc|}
                \hline
                Parameter&N&$x_1/x_2$&$\gamma$&Confidence\\
                \hline
                Stars (V-band)&&1.94&5/2&63\%\\
                AGN (V-band)&&10&5/2&97\%\\
                3CR (extragalactic)&298&8&5/2&96\%\\
                3CR (Galactic)&38&8&2&87\%\\
                4U  (Galactic)&181&18&2&94\%\\
                4U  (transients)&12&3.5&2&72\%\\
                SGR 1806$-$20 burst fluence&760&$7 \times 10^4$&1.7&99.96\%\\
                Crab Giant Pulses&$>1100$&1.25&2.8&33\%\\
                \hline
        \end{tabular}
        \end{center}
	\caption{\label{outliers1} Ratios of highest to second-highest
	fluxes and fluences $x_1/x_2$ from several astronomical catalogues.
	$N$, where applicable, is the number of objects in the catalogue,
	$\gamma$ is the fitted or theoretical differential slope, and
	Confidence is the confidence with which the hypothesis that
	$x_1/x_2$ is consistent with the power law can be rejected.
	References in \cite{K22c}.}
\end{table}

In each case in Table \ref{outliers1}, with one exception, the most extreme
object may be consistent with the power law fitted to the less extreme
objects, indicating that they are qualitatively similar.  The statistics
of variable sources (AGN, 3CR) are likely biased to higher extreme values
because the strongest sources are better observed, increasing the likelihood
of catching unusual excursions to yet higher flux; this may explain their
marginally significant disagreement with the extrapolated power laws.  The
one case in which the strongest event is clearly inconsistent with the
power law is the giant outburst of SGR 1806$-$20.  It must be qualitatively
different from its lesser outbursts.  This is usually attributed to a
global reorganization of the magnetic field, in contrast to lesser localized
flares.

\begin{table}
        \begin{center}
        \begin{tabular}{|crccc|}
                \hline
                Parameter&N&$x_1/x_2$&$\gamma$&Confidence\\
                \hline
                FRB Fluxes (Parkes)&31&4.3&5/2&89\%\\
                FRB Fluxes (UTMOST)&15&1.37&5/2&38\%\\
                FRB Fluxes (ASKAP)&42&1.15&5/2&19\%\\
                FRB Fluxes (CHIME)&536&1.33&2.4&33\%\\
                FRB Fluences (Parkes)&31&1.1&5/2&17\%\\
                FRB Fluences (UTMOST)&15&1.71&5/2&55\%\\
                FRB Fluences (ASKAP)&42&2.1&5/2&67\%\\
                FRB Fluences (CHIME)&536&1.0&2.4&0\%\\
                FRB 20200428 Fluxes&&$17000$&5/2&$>99.9999$\%\\
                FRB 20200428 Fluences&&$3600$&5/2&$>99.999$\%\\
                FRB RM&19&200&5/4&73\%\\
                FRB 20121102A Fluxes&93&1.55&1.7&26\%\\
                \hline
        \end{tabular}
        \caption{\label{outliers2} Ratios of most extreme to second
		most-extreme FRB.  Some $\gamma$ are for a Euclidean
		universe, $\gamma$ for RM is for a SNR model, others are
		fits.  Confidence is the confidence with which the
		hypothesis that $x_1/x_2$ is consistent with the power law
		can be rejected.  Entries for FRB 20200428 refer to its four
		observed bursts, not comparison to other FRB.  Details and
		references in \citet{K22c}.}
        \end{center}
\end{table} 

Analogous analyses for various FRB statistics are shown in Table
\ref{outliers2}.  The fluxes and fluences from various FRB catalogues (it is
necessary to analyze each catalogue separately so that the sample be
homogeneous) are all consistent with power laws, as are the outbursts of FRB
20121102A.  The exceptions are the flux and fluence of the giant outburst of
FRB 20200428, that are inconsistent, with very high statistical confidence,
with extrapolations of the power laws estimated from its three lesser
outbursts.  FRB 20200428 is consistent with expectations for a lower energy
analogue of the apparently non-repeating FRB at cosmological distances.
\subsection{Why No FRB Associated with SGR 1806$-$20}
The $10^{47}\,$ergs/s SGR 1806$-$20 filled its magnetosphere with dense
equilibrium pair plasma, preventing acceleration of energetic particles
and escape of radio radiation.  No particle acceleration or radio radiation
is expected from any SGR with (isotropic) power $\gtrsim 10^{42}\,$ergs/s.
At energy densities corresponding to pair equilibrium at temperatures
$k_B T > 22\,$keV, space on scales of a neutron star's inner magnetosphere
is filled with dense pair plasma \citep{K96}.  The threshold temperature is
a weak function of the length scale.  {The radiation from such a thermal
plasma would be unbeamed, unlike FRB that are likely beamed, an alternative
explanation.}

This argument implies that apparently non-repeating FRB are produced by
``intermediate'' outbursts of SGR, with total power $\lesssim
10^{42}\,$ergs/s.  It is unclear how these may differ from the Galactic
SGR with much more energetic giant outbursts.  For example, SGR that make
FRB might have {weaker} magnetic fields than those that produce giant
outbursts, or FRB may be produced by intermediate-scale outbursts of SGR
that are capable of (and may also produce) giant outbursts.  Galactic SGR
should be monitored for FRB activity.
\section{Distribution of FRB in the Universe}
CHIME/FRB has provided a homogeneous catalogue \citep{CHIME/FRB} of more
than 500 FRB sources, most apparently non-repeating.  This has enabled 
statistical studies that compare the distribution of FRB {\it vs.\/}
redshift $z$ to the distribution of other objects and processes,
specifically the star formation rate.  Several studies
\citep{ZZ22,Zhang24,Chen24} have found that the distributions differ, with
FRB concentrated at small $z$ while the star formation rate was much larger
in the more distant past (note, however, that \citet{WvL24} came to the
opposite conclusion).  Because the time lag from star formation to
neutron star formation is cosmologically brief (the minimum initial stellar
mass must be at least the Chandrasekhar mass $1.40 M_\odot$ and is likely
at least $6 M_\odot$, stars with main sequence lifetimes of $\sim 3 \times
10^9$ y and $\sim 10^8$ y, respectively) this argues against young neutron
stars as the sources of FRB.  Whether resembling ordinary pulsars, whose
duration of activity is $\sim 10^7$ y, or magnetars, whose duration of
activity is $\sim 10^4$ y, the additional delay is insignificant.  Old
neutron stars may undergo catastrophic events, such as double neutron star
mergers, after arbitrarily long delays, but these cannot explain repeating
FRB and are insufficient in number to explain apparent non-repeaters
\citep{H20}.

In contrast, the distribution on FRB in the universe is consistent with an
origin in black hole accretion disc funnels because black holes live
forever, gradually accreting mass.  Their number density only declines as
a result of cosmic expansion, not nearly as fast as the star formation rate.
\section{Conclusion}
FRB and SGR are the two kinds of astronomical events whose most extreme
members are true outliers, far exceeding extrapolations from lesser events.
There are few or no other examples of such distributions (the Sun far
exceeds other stars in apparent brightness, but this is a selection effect:
the Earth must be habitable).  This is evidence in support of the
association of apparently non-repeating FRB (which FRB 20200428 would be if
at a distance between about 100 kpc and 10 Mpc) with SGR.  The fact that
repeating FRB do not have {flux or fluence} outlier events confirms
their qualitative difference from apparent non-repeaters.

Some suggestions for observations follow:
\begin{itemize}
	\item Black hole accretion discs whose {angular mometum} axes
		might point to the Earth should be monitored for FRB or
		analogous activity, perhaps repeating.  These should include
		both stellar-mass (X-ray binary) and supermassive (AGN)
		accreting black holes.
	\item Galactic (and extra-Galactic) SGR should be monitored for
		FRB activity, plausibly producing rare strong FRB and also
		FRB weaker by orders of magnitude but not more frequent
		by orders of magnitude.  {The STARE2 \citep{B20b}
		and GReX \citep{C21} systems are sensitive to such events.}
\end{itemize}
\appendix
\section{Cutoff Frequency for Coherent Emission}
\label{cutoff}
A minimal hypothesis for coherent emission is that it has a cutoff frequency
equal to the plasma frequency at the Goldreich-Julian charge density
\begin{equation}
	\label{GJ}
	\rho_{GJ} = {{\vec B}\cdot{\vec \Omega} \over 2 \pi c},
\end{equation}
where $B$ is the surface magnetic field and $\Omega$ the angular rotation
rate of a pulsar.  The hypothesis is that coherent structures cannot exist
with fewer than ${\cal O}(1)$ electron per unit wavelength.  If the
electrons are moving relativistically then radiation from electrons
separated perpendicular to the direction of motion by many wavelengths may
radiate coherently, and the criterion is electrons per unit wavelength in
the direction of motion and per wavelength times the Lorentz factor in the
two transverse directions (X-ray laser undulators are an example).

The charge density of Eq.~\ref{GJ} corresponds to a plasma frequency, and
hypothesized cutoff frequency (taking ${\vec B}\cdot{\vec \Omega} =
B\Omega$),
\begin{equation}
	\label{nup}
	\nu_p = \sqrt{B \Omega e \over 2 \pi^2 m_e c}.
\end{equation}
Numerically, for the Crab pulsar $\nu_p \approx 30\,$GHz, for SGR
1935$+$2154 (FRB 20200428) $\nu_p \approx 10\,$GHz.  {These cutoffs only
apply near the stellar surface, where $B$ is maximum.  For a dipole field
$\nu_p$ falls off $\propto r^{-3/2}$.}  Nearly all millisecond
pulsars have $B/P > 3 \times 10^{10}\,$Gauss/s, where $P$ is the period,
corresponding to $\nu_p > 400\,$MHz; these objects are generally observed at
low frequencies.  Eq.~\ref{nup} is consistent with these observations, and
its predictions of cutoffs may be testable.  The few pulsars that appear to
violate this condition may be attributable to erroneously low $B$ inferred
when low apparent spindown is the result of acceleration by the Galactic
gravitational field cancelling a larger true spindown rate \citep{P19}.

Eq.~\ref{nup} is not applicable to the proposed accretion disc sources of
frequently repeating FRB because the disc is a source of comparatively high
density plasma, without requiring pair production as in pulsars.  Bursts
may be emitted in the vacuum of an accretion funnel {where the density
may be much less than the Goldreich-Julian value.  The observation of FRB
20180916B at frequencies as low as 110 MHz \citep{P21a} constrains the
plasma density along the line of sight to the emission region to be $n_e \ll
1.5 \times 10^8\,$cm$^{-3}$.  This is less than the density $n_e \sim
10^9\,$cm$^{-3}$ inferred from DM variations observed for FRB 20190520B
\citep{AT23} in C-band (4--8 GHz), but at those higher frequencies the
propagation condition is $n_e \ll 10^{10}\,$cm$^{-3}$.}
\section{Conditions for Particle Acceleration}
\label{conditions}
In order to accelerate energetic particles, necessary both for incoherent
emission (as in AGN and radio sources) and for coherent emission (as in 
pulsars and FRB, if these particles drive a plasma instability), it is
necessary that they gain energy from an electric field faster than they lose
it to interaction with ambient plasma (by ``Coulomb drag'') \citep{A09}.
For a relativistic electron the ratio of these quantities defines an
acceleration parameter
\begin{equation}
	\label{A}
	A \approx {E m_e c^2 \over 4 \pi e^3 n \ln{\Lambda}},
\end{equation}
where $E$ is the electric field, $n$ the plasma particle density and
$\Lambda$ is $2 m_e c^2/I$ with $I$ the ionization potential in a neutral
medium or $m_ec^2/\hbar \omega_p$ in a plasma.  $\ln{\Lambda} \approx 20$
in most astronomical environments, and is insensitive to their parameters.
$A > 1$ is a necessary condition for particle acceleration.
\subsection{Near-vacuum magnetospheres}
In a rotating pulsar or magnetar magnetosphere the particle density may be
very low, conducive to particle acceleration.  The minimum density is the
Goldreich-Julian density
\begin{equation}
	n = {{\vec \Omega}\cdot{\vec B} \over 2 \pi c e},
\end{equation}
where $\vec \Omega$ is the rotation rate and $\vec B$ the local magnetic
field.  Then, taking ${\vec \Omega} \cdot {\vec B} = \Omega B$ and $E =
\Omega R B/c$ at a radius $R$, the acceleration parameter Eq.~\ref{A}
\begin{equation}
	A = {E m_e c^3 \over 2 \ln{\Lambda} B e^2} \approx {m_e c^2 \over
	2 \ln{\Lambda} (e^2/R)}.
\end{equation}
Because $R$ is the overall system dimension, at least $\sim 10\,$km,
$e^2/R \lesssim 10^{-24}\,$erg and $A \gtrsim 10^{17}$.

Such magnetospheres are natural sites for acceleration of energetic
particles.  The passage of accelerated electrons through background plasma,
ambient or positrons created by pair production, may make excitation of
plasma waves and coherent radiation, such as pulsar and FRB emission,
almost unavoidable when there is no copious source (such as accretion) of
thermal plasma to put the plasma in a magnetohydrodynamic regime.

\end{document}